# The Gravitational Universe

A science theme addressed by the *eLISA* mission observing the entire Universe

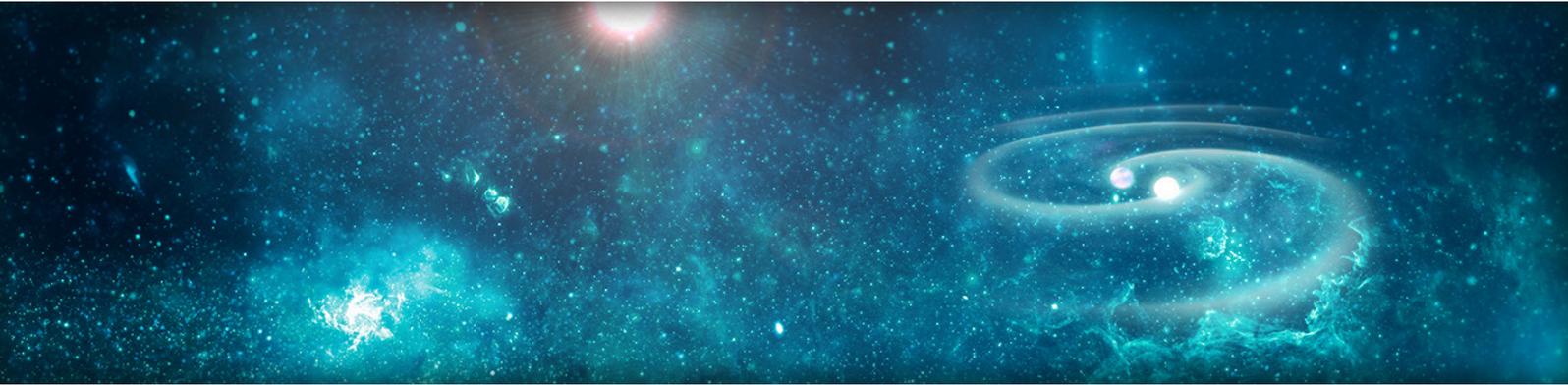

**Prof. Dr. Karsten Danzmann**
Albert Einstein Institute Hannover

MPI for Gravitational Physics and
Leibniz Universität Hannover
Callinstr. 38
30167 Hannover
Germany

karsten.danzmann@aei.mpg.de

Tel.: +49 511 762 2229
Fax: +49 511 762 2784

Detailed information at
http://elisascience.org/whitepaper

*The last century has seen enormous progress in our understanding of the Universe. We know the life cycles of stars, the structure of galaxies, the remnants of the big bang, and have a general understanding of how the Universe evolved. We have come remarkably far using electromagnetic radiation as our tool for observing the Universe. However, gravity is the engine behind many of the processes in the Universe, and much of its action is dark. Opening a gravitational window on the Universe will let us go further than any alternative. Gravity has its own messenger: Gravitational waves, ripples in the fabric of spacetime. They travel essentially undisturbed and let us peer deep into the formation of the first seed black holes, exploring redshifts as large as $z \sim 20$, prior to the epoch of cosmic re-ionisation. Exquisite and unprecedented measurements of black hole masses and spins will make it possible to trace the history of black holes across all stages of galaxy evolution, and at the same time constrain any deviation from the Kerr metric of General Relativity. eLISA will be the first ever mission to study the entire Universe with gravitational waves. eLISA is an all-sky monitor and will offer a wide view of a dynamic cosmos using gravitational waves as new and unique messengers to unveil The Gravitational Universe. It provides the closest ever view of the early processes at TeV energies, has guaranteed sources in the form of verification binaries in the Milky Way, and can probe the entire Universe, from its smallest scales around singularities and black holes, all the way to cosmological dimensions.*

# AUTHORS

Pau Amaro Seoane, Sofiane Aoudia, Gerard Auger, Stanislav Babak, Enrico Barausse, Massimo Bassan, Volker Beckmann, Pierre Binétruy, Johanna Bogenstahl, Camille Bonvin, Daniele Bortoluzzi, Julien Brossard, Iouri Bykov, Chiara Caprini, Antonella Cavalleri, Monica Colpi, Giuseppe Congedo, Karsten Danzmann, Luciano Di Fiore, Marc Diaz Aguilo, Ingo Diepholz, Rita Dolesi, Massimo Dotti, Carlos F. Sopuerta, Luigi Ferraioli, Valerio Ferroni, Noemi Finetti, Ewan Fitzsimons, Jonathan Gair, Domenico Giardini, Ferran Gibert, Catia Grimani, Paul Groot, Hubert Halloin, Gerhard Heinzel, Martin Hewitson, Allan Hornstrup, David Hoyland, Mauro Hueller, Philippe Jetzer, Nikolaos Karnesis, Christian Killow, Andrzej Krolak, Ivan Lloro, Davor Mance, Thomas Marsh, Ignacio Mateos, Lucio Mayer, Joseph Moerschell, Gijs Nelemans, Miquel Nofrarias, Frank Ohme, Michael Perreur-Lloyd, Antoine Petiteau, Eric Plagnol, Edward Porter, Pierre Prat, Jens Reiche, David Robertson, Elena Maria Rossi, Stephan Rosswog, Ashley Ruiter, B.S. Sathyaprakash, Bernard Schutz, Alberto Sesana, Benjamin Sheard, Ruggero Stanga, Tim Sumner, Takamitsu Tanaka, Michael Tröbs, Hai-Bo Tu, Daniele Vetrugno, Stefano Vitale, Marta Volonteri, Gudrun Wanner, Henry Ward, Peter Wass, William Joseph Weber, Peter Zweifel

*Affiliations can be found in a detailed list of authors at http://elisascience.org/authors*

# CONTRIBUTORS

Heather Audley, John Baker, Simon Barke, Matthew Benacquista, Peter L. Bender, Emanuele Berti, Nils Christopher Brause, Sasha Buchman, Jordan Camp, Massimo Cerdonio, Giacomo Ciani, John Conklin, Neil Cornish, Glenn de Vine, Dan DeBra, Marc Dewi Freitag, Germán Fernández Barranco, Filippo Galeazzi, Antonio Garcia, Oliver Gerberding, Lluis Gesa, Felipe Guzman Cervantes, Zoltan Haiman, Craig Hogan, Daniel Holz, C.D. Hoyle, Scott Hughes, Vicky Kalogera, Mukremin Kilic, William Klipstein, Evgenia Kochkina, Natalia Korsakova, Shane Larson, Maike Lieser, Tyson Littenberg, Jeffrey Livas, Piero Madau, Peiman Maghami, Christoph Mahrdt, David McClelland, Kirk McKenzie, Sean McWilliams, Stephen Merkowitz, Cole Miller, Shawn Mitryk, Soumya Mohanty, Anneke Monsky, Guido Mueller, Vitali Müller, Daniele Nicolodi, Samaya Nissanke, Kenji Numata, Markus Otto, E. Sterl Phinney, Scott Pollack, Alix Preston, Thomas Prince, Douglas Richstone, Louis Rubbo, Josep Sanjuan, Stephan Schlamminger, Daniel Schütze, Daniel Shaddock, Sweta Shah, Aaron Spector, Robert Spero, Robin Stebbins, Gunnar Stede, Frank Steier, Ke-Xun Sun, Andrew Sutton, David Tanner, Ira Thorpe, Massimo Tinto, Michele Vallisneri, Vinzenz Wand, Yan Wang, Brent Ware, Yinan Yu, Nicolas Yunes

*Affiliations can be found in a detailed list of contributors at http://elisascience.org/contributors*

# SUPPORTERS

Among the, roughly, 1000 scientific supporters of the Gravitational Universe science theme, are

GERARDUS 'T HOOFT *Utrecht University (Netherlands)*, BARRY BARISH *Caltech (United States)*, CLAUDE COHEN-TANNOUDJI *College de France (France)*, NEIL GEHRELS *NASA Goddard Space Flight Center (United States)*, GABRIELA GONZALEZ *LIGO Scientific Collaboration Spokesperson, LSU (United States)*, DOUGLAS GOUGH *Institute of Astronomy, University of Cambridge (United Kingdom)*, STEPHEN HAWKING *University of Cambridge, DAMTP (United Kingdom)*, STEVEN KAHN *Stanford University/SLAC National Accelerator Laboratory (United States)*, MARK KASEVICH *Stanford University, Physics Dept. (United States)*, MICHAEL KRAMER *Max-Planck-Institut fuer Radioastronomie (Germany)*, ABRAHAM LOEB *Harvard University (United States)*, PIERO MADAU *University of California, Santa Cruz (United States)*, LUCIANO MAIANI *Università di Roma La Sapienza (Italy)*, JOHN MATHER *NASA Goddard Space Flight Center (United States)*, DAVID MERRITT *Rochester Institute of Technology (United States)*, VIATCHESLAV MUKHANOV *LMU München (Germany)*, GIORGIO PARISI *Universita di Roma la Sapienza (Italy)*, STUART SHAPIRO *University of Illinois at Urbana-Champaign (United States)*, GEORGE SMOOT *Universite Paris Diderot (France)*, SAUL TEUKOLSKY *Cornell University (United States)*, KIP THORNE *California Institute of Technology (United States)*, GABRIELE VENEZIANO *Collège de (France) (France)*, JEAN-YVES VINET *Virgo Collaboration Spokesperson, OCA Nice (France)*, RAINER WEISS *MIT (United States)*, CLIFFORD WILL *University of Florida (United States)*, EDWARD WITTEN *Institute for Advanced Study, Princeton (United States)*, ARNOLD WOLFENDALE *Durham University (United Kingdom)*, and SHING-TUNG YAU *Harvard University (United States)*.

*A complete list of supporters can be found at http://elisascience.org/supporters*

# INTRODUCTION

In the early years of this millennium, our view of the Universe has been comfortably consolidated in some aspects, but also profoundly changed in others. In 2003, the double pulsar PSR J0737-3039 was discovered [1–2], and General Relativity passed all of the most stringent precision tests in the weak-field limit.

In 2013, the toughest test on General Relativity was performed through the observation of PSR J0348+0432, a tightly-orbiting pair of a newly discovered pulsar and its white-dwarf companion. Given the extreme conditions of this system, some scientists thought that Einstein's equations might not accurately predict the amount of gravitational radiation emitted, but General Relativity passed with flying colours [3]. However, no test of General Relativity could be considered complete without probing the strong-field regime of gravitational physics, where mass motions are close to the speed of light, $c$, and gravitational potentials close to $c^2$: Around a black hole, a central singularity protected by an event horizon, relativistic gravity is extreme. Exploring the physics of the inspiral of a small compact object skimming the event horizon of a large black hole, or the physics of a black hole-black hole collision, is probing gravity in the relativistic strong-field limit.

2013 marked an important date: ESA's Planck mission confirmed the Λ-CDM paradigm of cosmology at an unprecedented level of accuracy, offering the most precise all-sky image of the distribution of dark matter across the entire history of the Universe [4], further confirming that the minuscule quantum fluctuations which formed at the epoch of inflation were able to grow hierarchically from the small to the large scale under the effect of the gravity of dark matter, and eventually evolved into the galaxies we observe today with their billions of stars and central black holes [5]. In the Λ-CDM model, the first black holes, called seeds, formed in dark matter halos from the dissipative collapse of baryons [6–7], and as halos clustered and merged, so did their embedded black holes [8–9]. As a consequence, binary black holes invariably form, driven by galaxy collisions and mergers, and trace the evolution of cosmic structures.

In 2003 an ongoing merger between two hard X-ray galactic nuclei, with the characteristics of an Active Galactic Nucleus (AGN), was discovered just 150 Mpc away in the ultra-luminous infrared galaxy NGC 6240 (see Figure 1) [10]. Recent optical surveys have shown evidence of dual AGN [11] in today's Universe; even Andromeda and our Milky Way are due to collide and both house a central black hole!

Galaxy mergers were even more frequent in the past. The ensuing coalescence of massive black holes offers a new and unique tool, not only to test theories of gravity and the black hole hypothesis itself, but to explore the Universe from the onset of the cosmic dawn to the present.

In 2000, we discovered that dormant black holes are ubiquitous in nearby galaxies and that there are relationships between the black hole mass and the stellar mass of the host galaxy [12–13]. This gave rise to the concept that black holes and galaxies evolve jointly. Black holes trace galaxies and affect their evolution; likewise, galaxies trace black holes and affect their growth. Coalescing binary black holes pinpoint the places and times where galaxies merge, revealing physical details of their aggregation.

This new millennium also witnessed the discovery of numerous ultra-compact binary systems containing white dwarfs and/or neutron stars in close orbit, in the Milky Way [14]. These are excellent laboratories for exploring the extremes of stellar evolution in binary systems. They will transform into Type Ia supernovae or into merging binaries which will soon be detected by the ground-based gravitational wave detectors LIGO and VIRGO.

Stellar mass black holes with a pulsar as companion remain elusive, as they are very rare systems. Tracing these almost dark ultra-compact binaries of all flavours with gravitational waves all over the Milky Way will reveal how binary stars formed and evolved in the disc and halo of the Galaxy. Some of these are already known through electromagnetic observations and serve as guaranteed verification sources.

All of these advances were possible using only our first 'sense' for observing the Universe, electromagnetic radiation, tracing electromagnetic interactions of baryonic matter in the Universe. However, almost all of the Universe remains electromagnetically dark. On astronomical scales gravitation is the real engine of the Universe. By 'listening' to gravity we will be able to see further than ever before.

We can 'listen' to the Universe by directly observing gravitational waves, ripples in the fabric of spacetime travelling at the speed of light, which only weakly interact with matter and travel largely undisturbed over cosmological distances. Their signature is a fractional squeezing of spacetime perpendicuar to the direction of propagation, with an amplitude $h = \Delta L/L$ on the order of $10^{-20}$. Laser interferometry is a standard tool for such measurements and has been under development for over 30 years now [15].

Electromagnetic observations of the Universe, plus theoretical modelling, suggest that the richest part of the gravitational wave spectrum falls into the frequency range accessible to a space interferometer, from about 0.1 mHz to 100 mHz. In this band, important first-hand information can be gathered to tell us how binary stars formed in our Milky Way, and to test the history of the Universe out to redshifts of order $z \sim 20$, probing gravity in the dynamical strong-field regime and on the TeV energy scale of the early Universe [16]. *eLISA* will be the first ever mission to survey the entire Universe with gravitational waves, addressing the science theme *The Gravitational Universe*. The Next Gravitational Observatory (NGO) mission concept studied by ESA for the L1 mission selection is used as a strawman mission concept for *eLISA* [15]. ∎

> *In Sections I, II, and III, we will describe the eLISA observational capabilities in terms of an observatory sensitivity that is shown in Figure 12 as a U-shaped curve. The curve is based upon the sensitivity model of the strawman mission concept presented in Section IV. For the purpose of the discussion in Sections I, II, and III, this is just a baseline requirement. The question of how this requirement can be met, or exceeded, in an actual mission is addressed in Section IV.*

# I. ASTROPHYSICAL BLACK HOLES

As we will discuss in more detail in this section, *eLISA* observations will probe massive black holes over a wide, almost unexplored, range of redshift and mass, covering essentially all important epochs of their evolutionary history. *eLISA* probes are coalescing massive binary black holes, which are among the loudest sources of gravitational waves in the Universe. They are expected to appear at the 'cosmic dawn', around a redshift of $z \sim 11$ or more, when the first galaxies started to form. Coalescing binary black holes at a redshift as remote as $z \sim 20$ can be detected with *eLISA*, if they exist. *eLISA* will also explore black holes through 'cosmic high noon' (a term introduced by [17]), at redshifts of $z \sim 3$ to $z \sim 1.5$, when the star formation rate in the Universe and the activity of Quasi Stellar Objects (QSOs) and AGNs was highest. In the 'late cosmos', at $z < 1$, *eLISA* will continue to trace binary mergers, but it will also detect new sources, the Extreme Mass Ratio Inspirals (EMRIs), i.e., the slow inspiral and merger of stellar mass black holes into large, massive black holes at the centres of galaxies. These are excellent probes for investigating galactic nuclei during the AGN decline.

## I.I Massive binary black holes

The exploration of the sky across the electromagnetic spectrum has progressively revealed the Universe at the time of the cosmic dawn. The most distant star collapsing into a stellar mass black hole, the Gamma Ray Burst GRB 090429B, exploded when the Universe was 520 Myr old (at a redshift of $z \sim 9.4$), confirming that massive stars were born and died very early on in the life of the Universe [18]. The most distant known galaxy, MACS0647-JD, at a redshift of $z \sim 10.7$, was already in place when the Universe was about 420 Myr old [19], and ULAS J1120+0641 holds the record for being the most distant known QSO, thus the most distant supermassive ($\sim 10^9\,M_\odot$) accreting black hole at redshift $z \sim 7.08$, about 770 Myr after the big bang [20]. Such observations clearly show that stars, black holes, and galaxies, the key, ubiquitous components of the Universe, were present before the end of the reionisation phase around $z \sim 6$ [21]. These are the brightest sources, probing only the peak of an underlying distribution of smaller objects: the less luminous pre-galactic discs, and the less massive stars and black holes, of which little is known. Even the brightest QSOs fade away in the optical regime due to the presence of neutral hydrogen in the intergalactic medium (the Gunn-Peterson trough, [22]), and the search for the deepest sources using X-rays may be hindered by intrinsic obscuration, confusion due to crowding and the unresolved background light [23].

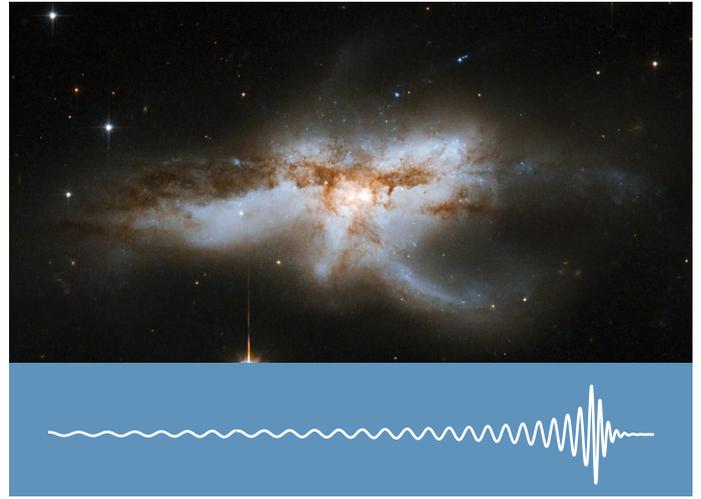

Figure 1: **Merging galaxy NGC 6240 and a representative waveform of the expected gravitational waves from the coalesence of two supermassive black holes.** Observations with NASA's Chandra X-ray observatory have disclosed two giant black holes inside NGC 6240. They will drift toward one another and eventually merge into a larger black hole. *Credit: NASA, ESA, the Hubble Heritage (STScI/AURA)-ESA/Hubble Collaboration, and A. Evans (University of Virginia, Charlottesville/NRAO/Stony Brook University).*

The entire zoo of objects, which in the past formed the small building blocks of the largest ones we see today, is so far pretty much unexplored. These primitive objects started to form at the onset of the cosmic dawn, around $z \sim 20$–$30$, according to current cosmological models [24]. In fact, simulations indicate that the very early pre-galactic, gas-rich discs had low masses, small luminosities and were very metal-poor. At an epoch of $z \sim 20$ to 30, the earliest stars may have had masses exceeding $100\,M_\odot$, ending their lives as comparable stellar mass black holes, providing the seeds that would later grow into supermassive black holes [6, 25]. However, as larger, more massive and metal enriched galactic discs progressivly formed, other paths for black hole seed formation became viable (see [26] for a review). Global gravitational instabilities in gaseous discs may have led to the formation of quasi-stars of $10^3$–$10^4\,M_\odot$ that later collapsed into seed black holes [7]. Further alternatives arise in the form of the collapse of massive stars formed in run-away stellar collisions in young, dense star clusters [27] or the collapse of unstable self-gravitating gas clouds in the nuclei of gas-rich galaxy mergers at later epochs [28]. Thus, the initial mass of seed black holes remains one of the largest uncertainties in the present theory of black hole formation, as the mechanism is still unknown, and the electromagnetic horizon too small for the direct detection of the formation of individual seeds.

Most of the investigations of galaxies and AGNs in the electromagnetic Universe, in terms of richness of sources, focus on a later epoch: the cosmic high noon, a period around $z \sim 1.5$–$3$. This epoch features several critical transformations in galaxy evolution, since around that time both the luminous QSOs and the star formation rate were at their peak [29–30]. Galaxy mergers and accretion along filaments during cosmic high noon were likely to be the driving force behind the processes of star formation, black hole fueling, and galaxy growth. This turned starforming discs into larger discs or quenched spheroidal systems hosting supermassive black holes of billions of solar masses [31–33]. In this framework, massive black hole binaries inevitably form in large numbers, over a variety of mass scales, driven by frequent galaxy mergers [8–9, 34]. Signs of galaxy mergers with dual black holes at wide separations (on the order of kpc) come from observations of dual AGNs in optical and X-ray surveys, while observations of binary black holes with sub-pc scale separations remain uncertain and only candidates exist at present [35]. Studies of the dynamics of black holes in merging galaxies have shown that black hole coalescences trace the merger of the dense baryonic cores better than the mergers of dark halos, as their dynamics are sensitive to gas and star content and feed-back [36–37].

Theoretical models developed in the context of the Λ-CDM paradigm [38–41] have been successful in reproducing properties of the observed evolution of galaxies and AGNs, such as the colour distribution of galaxies, the local mass density and mass function of supermassive black holes, and the QSO luminosity function at several wavelengths out to $z \sim 6$. Information about the underlying population of inactive, less massive and intrinsically fainter black holes, which grew through accretion and mergers across all cosmic epochs, is still lacking and difficult to gather.

*The Gravitational Universe* proposes a unique, new way to probe both cosmic dawn and high noon, to address a number of unanswered questions:

- *When did the first black holes form in pre-galactic halos, and what is their initial mass and spin?*
- *What is the mechanism of black hole formation in galactic nuclei, and how do black holes evolve over cosmic time due to accretion and mergers?*
- *What is the role of black hole mergers in galaxy formation?*

*eLISA* will study the evolution of merging massive black holes across cosmic ages, measuring their mass, spin and redshift over a wide, as yet unexplored, range. Black holes with masses between $10^4 M_\odot$ and $10^7 M_\odot$ will be detected by *eLISA*, exploring for the first time the low-mass end of the massive black hole population, at cosmic times as early as $z \sim 10$, and beyond.

### eLISA discovery domain

Coalescing black hole binaries enter the *eLISA* sensitivity band from the low frequency end, sweeping to higher frequencies as the inspiral gets faster and faster, as shown in Figure 13. Eventually they merge, with the formation of a common event horizon, followed by the ringdown phase during which residual deformation is radiated away and a rotating (Kerr) black hole remnant is formed. The waveform detected by *eLISA* is a measure of the amplitude of the strain in space as a function of time in the rest frame of the detector. This waveform carries information about the masses and spins of the two black holes prior to coalescence, the inclination of the binary plane relative to the line of sight, the luminosity distance and sky location, among other parameters [42]. Complete waveforms have been designed by combining Post Newtonian expansion waveforms for the early inspiral phase with an analytical description of the merger and ringdown phase, calibrated against highly accurate, fully general relativistic numerical simulations of black hole coalescence [43–44]. The first figure of merit of the *eLISA* performance is the signal-to-noise ratio (SNR) of a massive black hole binary coalescence with parameters in the relevant astrophysical range. Figure 2 shows *eLISA* SNRs for equal mass, non-spinning coalescing binaries. Here we compute the SNR as a function of the total mass, *M*, and of the redshift, *z*, averaging over all possible source sky locations and wave polarisations, assuming two-year observations. The plot highlights the extraordinary capabilities of the instrument in covering almost all of the mass-redshift parameter space needed to trace black hole evolution. Binaries with $10^4 M_\odot < M < 10^7 M_\odot$ can be detected out to $z \sim 20$ with an SNR $\geq 10$, if they exist. Figure 2 shows that virtually all massive black holes in the Universe were loud *eLISA* sources at some point in their evolution.

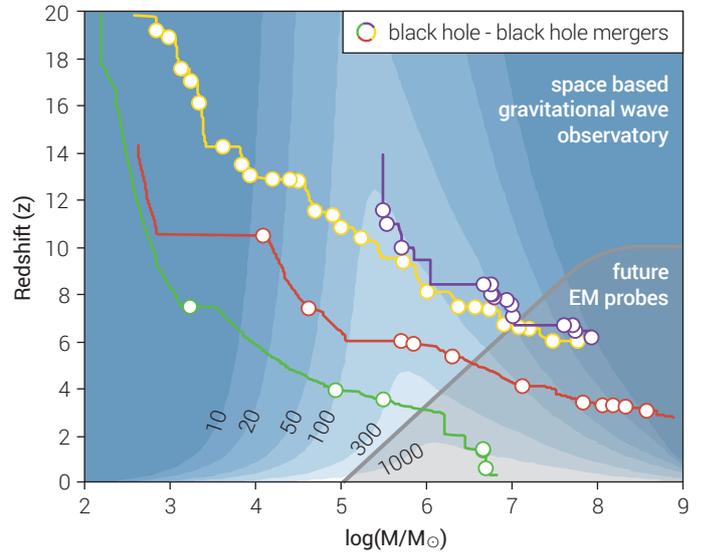

Figure 2: Constant-contour levels of the sky and polarisation angle-averaged SNR for eLISA, for equal mass non-spinning binaries as a function of their total rest frame mass, *M*, and cosmological redshift, *z*. The tracks represent the mass-redshift evolution of selected supermassive black holes: two possible evolutionary paths for a black hole powering a $z \sim 6$ QSO (starting from a massive seed, blue curve, or from a Pop III seed from a collapsed metal-free star, yellow curve); a typical $10^9 M_\odot$ black hole in a giant elliptical galaxy (red curve); and a Milky Way-like black hole (green curve). Circles mark black hole-black hole mergers occurring along the way. These were obtained using state of the art semi-analytical merger tree models [65]. The grey transparent area in the bottom right corner roughly identifies the parameter space for which massive black holes might power phenomena that will likely be observable by future electromagnetic probes.

Figure 3 shows error distributions in the source parameter estimation for events collected and extracted from a meta-catalogue of ~1500 simulated sources. The catalogue is constructed by combining predicted merger distributions from a number of cosmological models encompassing a broad range of plausible massive black hole evolution scenarios [45]. Uncertainties are evaluated using the Fisher Information Matrix approximation, which gives an estimate of the errors on the inferred parameters. Figure 3 illustrates that individual redshifted masses can be measured with unprecedented precision, with an error of 0.1% – 1% (we recall that observations can only determine the redshifted source masses, i.e., the product of mass and $(1+z)$). The spin of the primary black hole can be measured with very high accuracy, with 0.01 – 0.1 absolute uncertainty.

Current theoretical models predict coalescence rates in the range 10 – 100 per year [46–48]. For more than 10% of these, mostly occurring at a redshift of $z < 5$, the distance can be determined to better than a few percent and the sky location determined to better than a few degrees, which makes these sources suitable targets for coincident searches of electromagnetic counterparts (see Section V).

## Astrophysical impact

*eLISA* will allow us to survey the vast majority of all coalescing massive black hole binaries throughout the whole Universe. This will expose an unseen population of objects which will potentially carry precious information about the entire black hole population. It will provide both the widest and deepest survey of the sky ever, since gravitational wave detectors are essentially omni-directional by nature, and thus act as full-sky monitors. As highlighted in Figure 2, the range of black hole redshifts and masses that will be explored is complementary to the space explored by electromagnetic observations (see Figure 2).

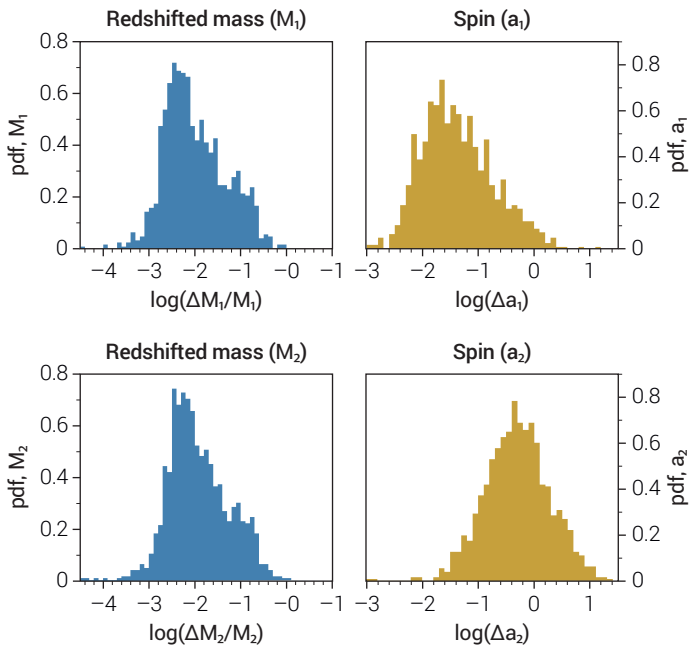

Figure 3: **eLISA parameter estimation accuracy for massive black hole binaries – probability density functions.** Left panels show errors on the redshifted mass, and right panels on the spins. Top panels refer to the primary black hole, bottom panels to the secondary black hole.

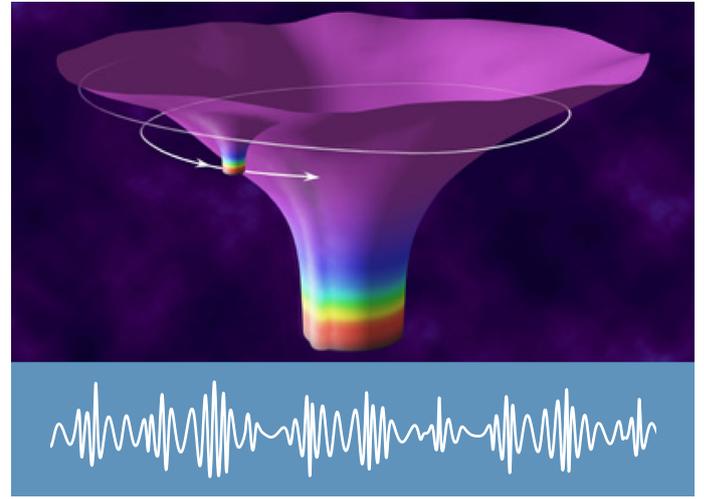

Figure 4: **An artist's impression of the spacetime of an extreme-mass-ratio inspiral and a representative waveform of the expected gravitational waves.** A smaller black hole orbits around a supermassive black hole. *Credit: NASA.*

*eLISA* will create the first catalogue of merging black holes, which will enable us to investigate the link between the growing seed population and the rich population of active supermassive black holes evolving during cosmic dawn and high noon. In doing this, we will probe the light end of the mass function at the largest redshifts and investigate the role of early black holes in cosmic re-ionisation and the heating of the intergalactic medium [49].

Black hole coalescence events will illuminate the physical process of black hole feeding. While the mass distribution carries information about the seeds, the spin distribution charts the properties of the accretion flows, whether they are chaotic or coherent [50]. Gravitational wave observations alone will be able to distinguish between the different evolution scenarios [46].

## I.II Extreme Mass Ratio Inspirals (EMRIs)

The present Universe is in a phase in which both the star formation rate and AGN activity are declining. In this late cosmos we observe quiescent massive black holes at the centres of galaxies within a volume of about 0.02 Gpc³ [51]. The current census comprises about 75 massive black holes out to $z < 0.03$. The black hole of $4 \times 10^6 \, M_\odot$ at the Galactic Centre is the most spectacular example [52–53]. Thanks to its proximity, a young stellar population has been revealed precisely where no young stars were predicted to form, as star-forming clouds are expected to be tidally disrupted there [54]. This indicates our lack of understanding about the origin of stellar populations around black holes, and in particular stellar dynamics, even in our own Galaxy. By probing the dynamics of intrinsically dark, relic stars in the nearest environs of a massive black hole, *eLISA* will offer the deepest view of galactic nuclei, exploring phenomena inaccessible to electromagnetic observations [55–56]. The probes used are the so-called EMRIs: a compact star (either a neutron star or a stellar mass black hole) captured in a highly relativistic orbit around the massive black hole and spiralling through the strongest field regions a few Schwarzschild radii from the event horizon before plunging into it (Figure 4).

Key questions can be addressed in the study of galactic nuclei with EMRIs:

- *What is the mass distribution of stellar remnants at the galactic centres and what is the role of mass segregation and relaxation in determining the nature of the stellar populations around the nuclear black holes in galaxies?*
- *Are massive black holes as light as $\sim 10^5 M_\odot$ inhabiting the cores of low mass galaxies? Are they seed black hole relics? What are their properties?*

*eLISA* will observe EMRI events, exploring the deepest regions of galactic nuclei, those near the horizons of black holes with masses close to the mass of the black hole at our Galactic Centre, out to redshifts as large as $z \sim 0.7$.

Stellar mass black holes are expected to dominate the observed EMRI rate for *eLISA*, as mass segregation by two-body relaxation tends to concentrate the heaviest stars near the central, massive black hole [16, 57–58], and a stellar mass black hole inspiral has a higher SNR, so it can be detected out to farther distances. EMRIs can be tracked around a central black hole for up to $10^4 – 10^5$ cycles on complex relativistic orbits (see Figure 5). As a consequence, the waveform carries an enormous amount of information [59]. Through observations of dark components alone, *eLISA* will detect EMRIs with an SNR > 20 in the mass interval for the central black hole between $10^4 M_\odot < M < 5 \times 10^6 M_\odot$ out to redshift $z \sim 0.7$ (see Figure 13), covering a co-moving volume of 70 Gpc³, a much larger volume than observations of dormant galactic nuclei today. The estimated detection rates, based on the best available models of the black hole population and of the EMRI rate per individual galaxy [60], are about 50 events for a 2 year mission, with a factor of 2 uncertainty from the waveform modelling and lack of knowledge about the system parameters, and an additional uncertainty of at least an order of magnitude stemming from the uncertain dynamics of dense stellar nuclei [61–62]. As shown in Figure 6, the masses of both black holes are, in most cases, measured to better than one part in $10^4$, the eccentricity at plunge is determined to a $10^{-4}$ accuracy, and the spin of the primary black hole to better than $10^{-3}$. The deviation of the quadrupole moment of the massive black hole with respect to the Kerr metric value is determined to better than 0.01, enabling unprecedented tests of General Relativity to be performed (see Section II).

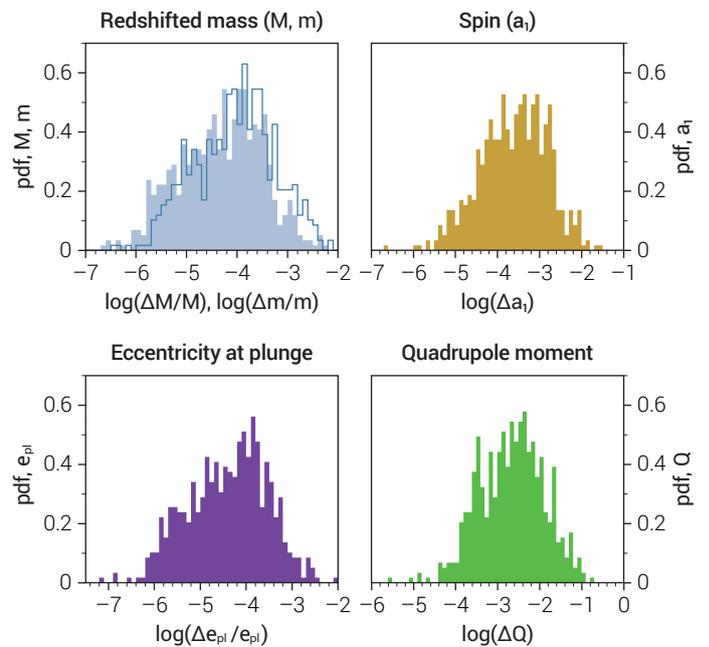

Figure 6: **eLISA parameter estimation accuracy for EMRIs.** Top left panel: estimation of the redshifted masses (filled: massive black hole; solid line: stellar black hole); top right panel: spin of the massive black hole; bottom left panel: eccentricity at plunge; bottom right panel: minimum measurable deviation of the quadrupole moment of the massive black hole from the Kerr value.

### Astrophysical impact

In *The Gravitational Universe*, EMRIs are exquisite probes for testing stellar mass black hole populations in galactic nuclei. With *eLISA* we will learn about the mass spectrum of stellar mass black holes, which is largely unconstrained both theoretically and observationally. The measurement of even a few EMRIs will give astrophysicists a totally new way of probing dense stellar systems, allowing us to determine the mechanisms that govern stellar dynamics in the galactic nuclei [58].

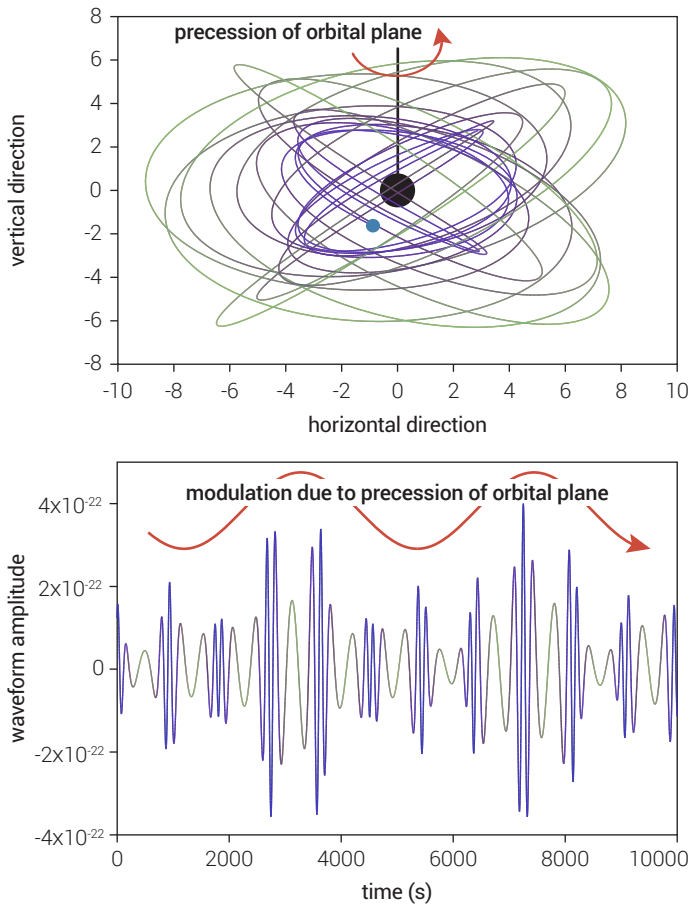

Figure 5: **EMRI orbit and signal.** In the top panel we see the geometrical shape of the ornate relativistic EMRI orbit. The lower panel shows the corresponding gravitational wave amplitude as a function of time.

Measurements of a handful of events will suffice to constrain the low end of the massive black hole mass function, in an interval of masses where electromagnetic observations are poor, incomplete or even missing [63]. By 2028 the Large Synoptic Survey Telescope (LSST) will have observed a large number of tidal disruption events [64], which will also teach us a lot about black holes and stellar populations in galactic centres. However, these events will typically be on the higher end of the black hole's mass function, and will not reveal masses with the same precision as *eLISA*, whose observations will give us information about the nature and the occupation fraction of massive black holes in low mass galaxies (as yet unconstrained) [65]. This will provide additional information about the origin of massive black holes, complementary to that gathered via observations of high $z$ massive black hole mergers. EMRIs will also provide the most precise measurements of Milky Way type massive black hole spins, and make it possible for us to investigate the spin distribution of single, massive black holes up to $z \sim 0.7$. EMRIs can occur around black holes in all galaxies, regardless of their nature, i.e., whether they are active or not. As such, spin measurements will not be affected by observational uncertainties in the spectra of the AGN.

Detection rates will tell us about the density of stellar mass black holes in the vicinity of the central black hole, constraining the effectiveness of the mass segregation processes [66]. The measured eccentricity and orbital inclination distributions can be linked directly to the preferred channel of EMRI formation, giving us important clues about the efficiency of dynamical relaxation, and the frequency of binary formation and breakup in dense nuclei [67]. ∎

## II. THE LAWS OF NATURE

This section covers tests of strong gravity and cosmology, two regimes where *The Gravitational Universe* will offer major advances.

### II.I High precision measurements of strong gravity

Einstein's General Relativity is one of the pillars of modern cosmology. The beauty of General Relativity is that it is a falsifiable theory: once the underlying mass distribution is identified to be a black hole binary system with fixed masses and spins, the theory has no further adjustable parameters. Thus even a single experiment incompatible with a prediction of the theory would lead to its invalidation, at least in the physical regime of applicability of the experiment.

*The Gravitational Universe* will explore relativistic gravity in the strong-field, non-linear regime. It seems unlikely that any other methods will achieve the sensitivity of *eLISA* to deviations of strong-field gravity by 2028 (see Section V). Unlike the ground-based instruments, *eLISA* will have sufficient sensitivity to observe even small corrections to Einstein gravity.

The strong-field realm of gravity theories can be probed near the event horizon of Kerr black holes or in other large-curvature environments (e.g., in the early Universe). Gravity can be thought of as strong in the sense that gravitational potentials are a significant fraction of $c^2$ or in the sense that the curvature tensor (or tidal force) is of very large magnitude. Testing strong gravity takes on different meanings depending upon which notion of 'strong gravity' is being used. In any case, both are very important to test with high precision. The measurement of the anisotropy of the Cosmic Microwave Background (CMB) by ESA's Planck satellite directly constrains properties of the quantum fluctuations during inflation which are thought to be the seeds of structure formation. This connection between physics on the smallest and largest scales is strong motivation for testing General Relativity to the highest possible accuracy, and in particular in the strong-field regime, where deviations could hint at a formulation of a quantum theory of gravity.

The nature of gravity in the strong-field limit is, so far, largely unconstrained, leaving open several questions:

- *Does gravity travel at the speed of light?*
- *Does the graviton have mass?*
- *How does gravitational information propagate: Are there more than two transverse modes of propagation?*
- *Does gravity couple to other dynamical fields, such as, massless or massive scalars?*
- *What is the structure of spacetime just outside astrophysical black holes? Do their spacetimes have horizons?*
- *Are astrophysical black holes fully described by the Kerr metric, as predicted by General Relativity?*

An outstanding way to answer these questions and learn about the fundamental nature of gravity is by observing the vibrations of the fabric of spacetime itself, for which coalescing binary black holes and EMRIs are ideal probes.

Exploring relativistic gravity with binary black hole mergers in the strong-field, dynamical sector

The coalescence of a massive black hole binary with mass ratio above one tenth generates a gravitational wave signal strong enough to allow detection of tiny deviations from the predictions of General Relativity. The signal comprises three parts—inspiral, merger and ringdown—each of which probes strong-field gravity. The inspiral phase is well understood theoretically: It can last several months in-band, and it could be observed with an SNR of tens to hundreds by *eLISA*. The non-linear structure of General Relativity, and possible deviations from it, are encoded in the phase and amplitude of the gravitational waves. Any effect that leads to a cumulative dephasing of a significant fraction of a wave cycle over the inspiral phase can be de-tected through matched filtering. Thousands of inspiral wave cycles will be observed, making it possible to detect even very small deviations in the inspiral rate predicted by General Relativity [68–71]. The propagation of gravitational waves can be probed through the dispersion of the inspiral signal. In General Relativity, gravitational waves travel with the speed of light (the graviton is massless) and interact very weakly with matter. Alternative theories with a massive graviton predict an additional frequency dependent phase shift of the observed waveform due to dispersion that depends on the graviton's mass, $m_g$, and the distance to the binary. An *eLISA*-like detector could set a bound around $m_g < 4 \times 10^{-30}$ eV [72], improving current Solar System bounds on the graviton mass, $m_g < 4 \times 10^{-22}$ eV, by several orders of magnitude. Statistical analysis of an ensemble of observations of black hole coalescences could also be used to place stringent constraints on theories with an evolving gravitational constant [73] and theories with Lorentz-violating modifications to General Relativity [74].

The inspiral is followed by a dynamical coalescence that produces a burst of gravitational waves. This is a brief event, comprising a few cycles and lasting about $5 \times 10^3$ sec $(M/10^6 M_\odot)$, yet it is very energetic, releasing $10^{59} (M/10^6 M_\odot)$ ergs of energy, corresponding to $10^{22}$ times the power of the Sun.

After the merger, the asymmetric remnant black hole settles down to a stationary and axisymmetric state through the emission of quasi-normal mode (QNM) radiation. In General Relativity, astrophysical black holes are expected to be described by the Kerr metric and characterised by only two parameters: mass and spin (the 'no-hair' theorem). Each QNM is an exponentially damped sinusoid with a characteristic frequency and damping time that depends only on these two parameters [75–77]. A measurement of two QNMs will therefore provide a strong-field verification that the final massive object is consistent with being a Kerr black hole [78–79]. The QNM spectrum of a black hole also has unique features which allow it to be distinguished from other (exotic) compact objects [80–83]. *eLISA* will observe ringdown signals with sufficient SNR to carry out these tests [84–86].

Exploring relativistic gravity with EMRIs in the strong-field, stationary, non-linear regime

EMRIs will provide a precise tool to probe the structure of spacetime surrounding massive black holes. The inspiralling compact object can generate hundreds of thousands of gravitational wave cycles while it is within ten Schwarzschild radii of the central black hole. These waveform cycles trace the orbit that the object follows, which in turn maps out details of the underlying spacetime structure, in a way similar to how stellar orbits have been used to precisely characterise the supermassive black hole at the centre of the Milky Way [87–88].

As seen in Figure 6, EMRI observations will not only provide very precise measurements of the 'standard' parameters of the system, but will provide strong constraints on departures of the central massive object from the Kerr black hole of General Relativity [89]. The no-hair theorem tells us that the stationary axisymmetric spacetime around it should be completely determined by its mass and spin parameter. The gravitational wave signal from an EMRI occurring in a 'bumpy' black hole spacetime in which the multipole moments differ from their Kerr values would show distinctive, detectable signatures [89–94]. Figure 6 shows that 10 % deviations in the mass quadrupole moment, $Q$, from the Kerr value would be detectable for any EMRI observed with an SNR greater than 20. For typical systems, 0.1 % deviations will be detectable, and for the best systems, 0.01 % deviations will be detectable [59].

An observed inconsistency with the Kerr multipole structure might indicate a surprisingly strong environmental perturbation, the discovery of a new type of exotic compact object consistent with General Relativity, or a failure in General Relativity itself, but these possibilities will be observationally distinguishable (for a review of different hypotheses see [95–97]). These deviations could exhibit themselves in the following ways: For a boson star, the EMRI signal would not shut off after the last stable orbit [98]. For horizonless objects such as gravastars, the orbiting body would resonantly excite the modes of the (putative) membrane replacing the black hole horizon [99], and for certain non-Kerr axisymmetric geometries, orbits could become ergodic [100] or experience extended resonances [101].

Alternatives that will be testable with *eLISA* observations include the dynamical Chern-Simons theory [102–105], scalar-tensor theories (with observable effects in neutron star-black hole systems where the Neutron Star (NS) carries scalar charge [106]), Randall-Sundrum-inspired braneworld models [107–108] and theories with axions that give rise to 'floating orbits' [109]. Generic alternatives could also be constrained using phenomenologically parametrised models [110].

Cosmography

*The Gravitational Universe* will use black hole binary mergers as 'standard sirens' to extract information on the expansion of the Universe, by measuring the expansion history with completely different techniques to electromagnetic probes. The term standard siren for gravitational wave sources, refers to a source that has its absolute luminosity encoded in its signal shape, analogous to a standard candle (like a Type Ia supernova) for electromagnetic sources. Black hole coalescences could serve as standard sirens for cosmography [111–112] by providing absolute and direct measurements of the luminosity distance, $D_L(z)$. When coupled with independent measurements of redshift, $z$, (for example, from associated transient electromagnetic sources ), these standard siren sources put points on the distance vs. redshift curve, and directly constrain the evolution history of the Universe. Several mechanisms have been proposed that will provide an electromagnetic counterpart to massive black hole coalescences detectable by *eLISA* [113], however, confident identification might be viable only for low redshift events. Although rare, events at redshift $z \lesssim 1-2$ are so loud that the baseline *eLISA* mission has the capability of localising them to within 10 square degrees, perhaps even 1 square degree, when information from the late merger of the black holes is included in the measurement model. This pins down these events on the sky well enough to allow searches for electromagnetic counterparts to the merger using wide area surveys such as LSST that will be active in 2028. With an associated counterpart, *eLISA* observations will allow 1 % measurements of $D_L(z)$ for 60 % of the sources, offering the prospect of ultra-precise determination of points on the distance-redshift curve that are completely independent of all existing constraints from Type Ia supernova, the CMB, etc. It is to be emphasised that there is no distance ladder in these measurements, since the luminosity distance is measured directly. This is possible because these sources are fundamentally understood, and General Relativity calibrates the distances. Weaker statistical constraints could also be derived in the absence of electromagnetic counterparts, using mergers at low redshift ($z < 2$) [114] or EMRIs [115–116]. EMRI observations could provide an independent measurement of $H_0$ to a precision of a few percent.

Impact on science

*The Gravitational Universe* will permit unprecedented measurements of General Relativity in the strong-field regime. *eLISA* will map the spacetime around astrophysical black holes, yielding a battery of precision tests of General Relativity in an entirely new regime. These have the potential to uncover hints about the nature of quantum gravity, as well as enabling measurements of the properties of the Universe on the largest scales.

## II.II Cosmology on the TeV energy scale — a fossil background of gravitational waves

Several processes occurring at very high energies in the primordial Universe can produce a stochastic background of gravitational waves. The detection of this relic radiation would have a profound impact both on cosmology and on high energy physics. Any fossil radiation of gravitational waves, if not washed away by inflation and later phase transitions, would have decoupled from matter and energy at the Planck scale. It can therefore directly probe cosmological epochs before the decoupling of the cosmic microwave background, currently our closest view of the big bang (Figure 7). The characteristic frequency of the gravitational waves is set by the horizon scale and therefore by the temperature of the Universe at the time of production. The *eLISA* frequency band of 0.1 mHz to 100 mHz corresponds to 0.1 to 100 TeV energy scales in the early Universe, at which new physics is expected to become visible. The

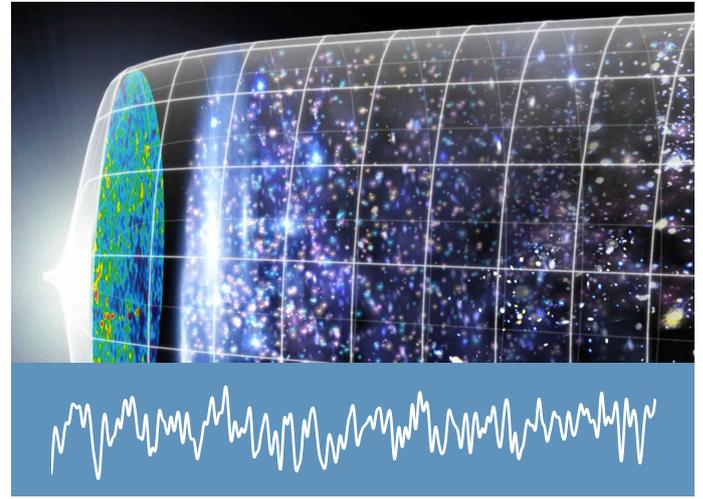

Figure 7: Evolution timeline (big bang and the early Universe) and a typical random waveform of the expected gravitational waves. Gravitational waves are the only way to see beyond the cosmic microwave background. Credit: NASA / WMAP science team.

Large Hadron Collider (LHC) has been built to investigate the physics operating at this energy scale, and in 2012 the experiment produced with the remarkable discovery of the Higgs boson which has completed the particle spectrum of the Standard Model. It is the final confirmation that spontaneous symmetry breaking is the mechanism at play in electroweak physics, and is the first example of a fundamental scalar field playing a role in a phase transition that took place in the very early Universe. These findings further motivate the search for a cosmic background of gravitational waves. *eLISA* would have the sensitivity to detect a relic background created by new physics active at TeV energies if more than a modest fraction, $\Omega_{GW} \sim 10^{-5}$ of the energy density of the Universe is, converted to gravitational radiation at the time of production.

A gravitational wave detector in space has the potential to revolutionise our understanding of the physics of the infant Universe by exploring the microphysical behaviour of matter and energy through the direct detection of gravitational waves produced at this epoch, rather than by observing collisions of elementary particles.

Discovery space

Abundant evidence suggests that the physical vacuum has not always been in its current state, and in many theories beyond the Standard Model, the conversion between vacuum states corresponds to a first-order phase transition. As the Universe expands and its temperature drops below the critical temperature, bubbles of a new phase form, expand, and collide, generating relativistic bulk flows, whose energy then dissipates in a turbulent cascade. The corresponding acceleration of matter radiates gravitational waves on a scale not far below the horizon scale [117–120]. *eLISA* could detect these gravitational waves, thus probing the Higgs field self couplings and potential, and the possible presence of supersymmetry, or of conformal dynamics at TeV scales. In general, since the Hubble length at the TeV scale is about 1 mm, the current threshold at which the effects of extra dimensions might appear happens to be about the same for experimental gravity in the laboratory and for the cosmological regime accessible to *eLISA*, thus allowing *eLISA* to probe the dynamics of warped sub-millimetre extra dimensions, present in the context of some string theory scenarios [121–122].

In some braneworld scenarios the Planck scale itself is not far above the TeV scale. Consequently, the reheating temperature would be in the TeV range and *eLISA* could probe inflationary reheating. After inflation the internal potential energy of the inflaton is converted into a thermal mix of relativistic particles, which can generate gravitational waves with an energy of about $10^{-3}$ or more of the total energy density[123–125]. *eLISA* could also probe gravitational waves produced directly by the amplification of quantum vacuum fluctuations during inflation, in the context of some unconventional inflationary models, such as pre-big bang or bouncing brane scenarios[126–128].

Phase transitions often lead to the formation of one-dimensional topological defects known as cosmic strings. Fundamental strings also arise as objects in string theory and, although formed on submicroscopic scales, it has been realised that these strings could be stretched to astronomical size by cosmic expansion [129–130]. Cosmic strings interact and form loops which decay into gravitationl waves; *eLISA* will be the most sensitive probe for these objects, offering the possibility of detecting direct evidence of fundamental strings. The spectrum from cosmic strings is distinguishably different from that of phase transitions or any other predicted source [130]: It has nearly constant energy per logarithmic frequency interval over many decades at high frequencies, offering the possibility of simultaneous detection by *eLISA* and ground-based interferometers. Moreover, if strings are not too light, occasional distinctive gravitational wave bursts might be observed from kinks or cusps on string loops. If detected, these individual bursts will provide irrefutable evidence for a cosmic string source. ∎

## III. ULTRA-COMPACT BINARIES IN THE MILKY WAY

Only a minority of the stars in the Universe are companionless, the majority being part of binary or multiple star systems (Figure 8). About half of the binaries form with sufficiently small orbital separations to interact and evolve into compact systems, often with white dwarfs, neutron stars, or possibly stellar mass black holes as components. The shortest period systems, known as ultra-compact binaries, are important sources of gravitational waves in the mHz frequency range [131]. These binaries are the outcome of one or more common-envelope phases that occur when one star evolves to the giant or supergiant stage. Unstable mass exchange leads to a short-lived phase in which the stellar core and the companion spiral towards each other, transferring angular momentum to the extended envelope of the giant that is blown away, leaving a tight binary system behind. This a necessary step in the formation of X-ray binaries, binary pulsars and double white dwarf binaries, observed in a variety of states and configurations [132]. Although the Milky Way is full of these sources, only a tiny fraction are currently observed and studied in-depth through observations from radio to X-rays.

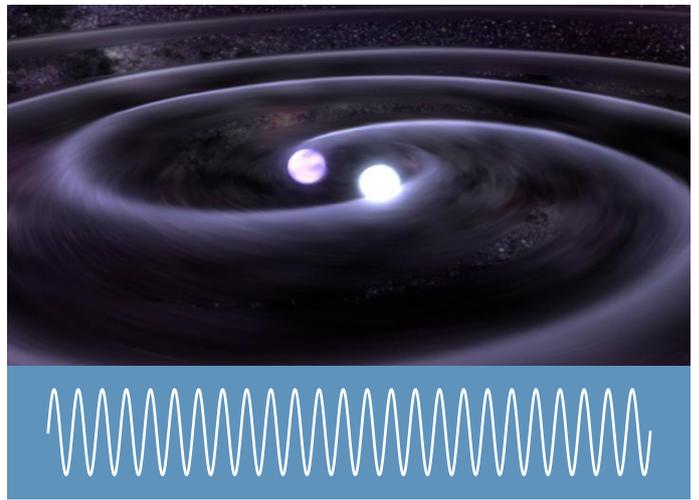

**Figure 8: Illustration of a compact binary star system and a representative waveform of the expected gravitational waves.** Two stars orbiting each other in a death grip are destined to merge all the while flooding space with gravitational waves. *Credit: GSFC/D.Berry.*

After the common-envelope phase, the compact stars in the binary are well separated, but evolve with shorter and shorter orbital periods due to the angular momentum loss via gravitational waves, until eventually the stars undergo another mass exchange once they reach orbital periods of minutes or less. Depending on the nature of the compact objects, they either merge or survive. Double neutron star binaries ultimately merge in bursts of high-frequency gravitational waves, potentially emitting a burst of electromagnetic radiation in the form of a short Gamma-Ray Burst. Although predicted to exist, no neutron star-stellar mass black hole binary or black hole-black hole binary has yet been detected. For binaries with a white dwarf component, mass loss is delicately balanced against loss of angular momentum, the outcome of which is unclear.

Almost always, the loss of energy through gravitational waves is so strong that it causes the system to merge and possibly to end in a Type Ia or sub-luminous supernova explosion [133–134] or in (rapidly spinning) neutron stars that may have millisecond radio pulsar or magnetar properties [135]. In the remaining small fraction of cases, the mass transfer stabilises the system and long-lived interacting binaries are formed (as in the AM CVn systems if the companion is a white dwarf, or ultra-compact X-ray binaries if it is a neutron star). The physics of this evolutionary junction is rich and diverse, it involves tides, mass transfer, highly super-Eddington accretion, and mass ejection. *eLISA* will provide the data necessary for us to quantify the roles that these various physical processes play.

Currently fewer than 50 ultra-compact binaries are known, only two of which have periods less than 10 minutes [14]; *eLISA* will discover several thousand of these. These systems are relatively short-lived and electromagnetically faint, but several known verification binaries are strong enough gravitational wave sources that they will be detected within weeks by *eLISA*. The discovery of many new ultra-compact binaries is one of the main objectives of *eLISA*, as it will provide a quantitative and homogeneous study of their populations and the astrophysics governing their formation. It also adds an additional facet to the knowledge of the Milky Way's structure: The distribution of sources in the thin disc, thick disc, halo and its globular clusters (known breeding grounds for the formation of compact binaries via dynamical exchanges) [136]. These detections will enable us to address a number of key questions:

- *How many ultra-compact binaries exist in the Milky Way?*
- *What is the merger rate of white dwarfs, neutron stars and stellar mass black holes in the Milky Way (thus better constraining the rate of the explosive events associated with these sources)?*
- *What does that imply for, or how does that compare to, their merger rates in the Universe?*
- *What happens at the moment a white dwarf starts mass exchange with another white dwarf or neutron star, and what does it tell us about the explosion mechanism of type Ia supernovae?*
- *What is the spatial distribution of ultra-compact binaries, and what can we learn about the structure of the Milky Way as a whole?*

To answer these questions *eLISA* will observe thousands of individual sources and for the first time capture the signal from a foreground of sources, the *sound* of millions of tight binaries.

### Discovery space

The vast majority of ultra-compact binaries will form an unresolved foreground signal in *eLISA* [137] as shown in Figure 13. Its average level is comparable to the instrument noise, but due to its strong modulation during the year (by more than a factor of two) it can be detected. The overall strength can be used to learn about the distribution of the sources in the Galaxy, as the different Galactic components (thin disc, thick disc, halo) contribute differently to the modulation [138]. Their relative amplitudes can be used to set upper limits on the, as yet completely unknown, halo population [139].

For a two-year *eLISA* mission, several thousand binaries are expected to be detected individually with an SNR > 7 [140] and their periods (below one hour and typically 5 – 10 min) determined from the periodicity of the gravitational wave signal. For many systems it will be possible to measure the first time derivative of the frequency, and thus determine the chirp mass (a combination of the masses of the two stars that can be used to distinguish white dwarf, neutron star and black hole binaries) and the distance to the source. For more than 100 sources concentrated around the inner Galaxy, we expect distance estimates with accuracies better than 1 %, enabling us to make a direct measurement of the distance to the Galactic Centre.

The number of ultra-compact binaries with neutron star or black hole components is still highly uncertain [141]. With *eLISA* operating as an all sky monitor, these systems can be observed throughout the Milky Way, providing a complete sample of binaries at the shortest periods (below 30 minutes), including ones containing stellar mass black holes, if they exist. The number of sources detected at mHz frequencies is directly related, via the gravitational wave orbital decay time scale, both to the number of systems formed at larger separations, and to the number of mergers. Therefore, the thousands of *eLISA* detections will also probe the formation of these binaries and their coalescence rates. Using *eLISA*, the sky position of about fifteen hundred sources will be determined to better than ten square degrees and more than half will have their distance determined to better than 20 % [142]. Several hundred of these may be found with current and future wide-field instruments such as the Visible and Infrared Survey Telescope for Astronomy (VIRCAM) and LSST. A few dozen of the highest SNR sources will have error boxes that are significantly smaller and can be realistically found with small field of view (several arcmin) cameras such as the Multi-adaptive Optics Imaging Camera for Deep Observations (MICADO) on the European Extremely Large Telescope (E-ELT) or the James Webb Space Telescope (JWST).

### Astrophysical impact

The large number of ultra-compact binaries discovered, and the fact that the sample is complete at the shortest periods, will help determine the total number of systems of all types as well as their merger rates. The systems containing neutron stars and/or black holes will be observed about a million years prior to coalescence, a phase that is still unexplored.

The highest SNR systems will allow a study of the complex physics of white dwarf mergers or of how systems survive as interacting binaries. Recent detailed simulations [143] have cast doubt on the theory that the actual merger would be a truly dynamical process taking only one or two orbits, and instead show that the merger would take place over many orbits, possibly allowing *eLISA* to observe some mergers directly.

*eLISA* will detect ultra-compact binaries beyond the Galactic Centre and in the Milky Way's halo using observations which are unaffected by dust obscuration, providing an independent probe of the components and formation history of the Milky Way. ∎

# IV. A STRAWMAN MISSION FOR THE eLISA SPACE GRAVITATIONAL WAVE OBSERVATORY

All of the above scientific objectives can be addressed by a single L-class mission consisting of 3 drag-free spacecraft forming a triangular constellation with arm lengths of one million km and laser interferometry between "free-falling" test masses. The interferometers measure the variations in light travel time along the arms due to the tidal deformation of spacetime by gravitational waves. Compared to the Earth-based gravitational wave observatories like LIGO and VIRGO, *eLISA* addresses the much richer frequency range between 0.1 mHz and 1 Hz, which is inaccessible on Earth due to arm length limitations and terrestrial gravity gradient noise.

The Next Gravitational wave Observatory (NGO) mission studied for the L1 selection [15] is an *eLISA* strawman mission concept. It enables the ambitious science program described here, and has been evaluated by ESA as both technically feasible and compatible with the L2 cost target. Its foundation is mature and solid, based on decades of development for LISA, including a mission formulation study, and the extensive heritage of flight hardware and ground preparation for the upcoming LISA Pathfinder geodesic explorer mission, which will directly test most of the *eLISA* performance and validate the *eLISA* instrumental noise model [144–145].

## Mission design

The NGO mission has three spacecraft, one 'mother' at the vertex and two 'daughters' at the ends, which form a single Michelson interferometer configuration (Figure 9). The spacecraft follow independent heliocentric orbits without any station-keeping and form a nearly equilateral triangle in a plane that is inclined by 60° to the ecliptic. The constellation follows the Earth at a distance between 10° and 30°, as shown in Figure 10. Celestial mechanics causes the triangle to rotate almost rigidly about its centre as it orbits around the sun, with variations of arm length and opening angle at the percent level.

The payload consists of four identical units, two on the mother spacecraft and one on each daughter spacecraft (Figure 11). Each unit contains a Gravitational Reference Sensor (GRS) with an embedded free-falling test mass that acts both as the end point of the optical length measurement, and as a geodesic reference test particle. A telescope with 20 cm diameter transmits light from a 2 W laser at 1064 nm along the arm and also receives a small fraction of the light sent from the far spacecraft. Laser interferometry is performed on an optical bench placed between the telescope and the GRS.

On the optical bench, the received light from the distant spacecraft is interfered with the local laser source to pro-duce a heterodyne beat note signal between 5 and 25 MHz, which is detected by a quadrant photodiode. The phase of that beat note is measured with μcycle/√Hz precision by an electronic phasemeter. Its time evolution reflects the laser light Doppler shift from the relative motion of the spacecraft, and contains both the macroscopic arm length variations on times cales of months to years, and the small fluctuations with periods between seconds and hours that represent the gravitational wave science signal. The measurement of relative spacecraft motion is then summed with a similar local interferometer measurement of the displacement between test mass and spacecraft. This yields the desired science measurement between distant free-falling test masses, removing the much larger motion of the spacecraft, which contains both thruster and solar radiation pressure noise.

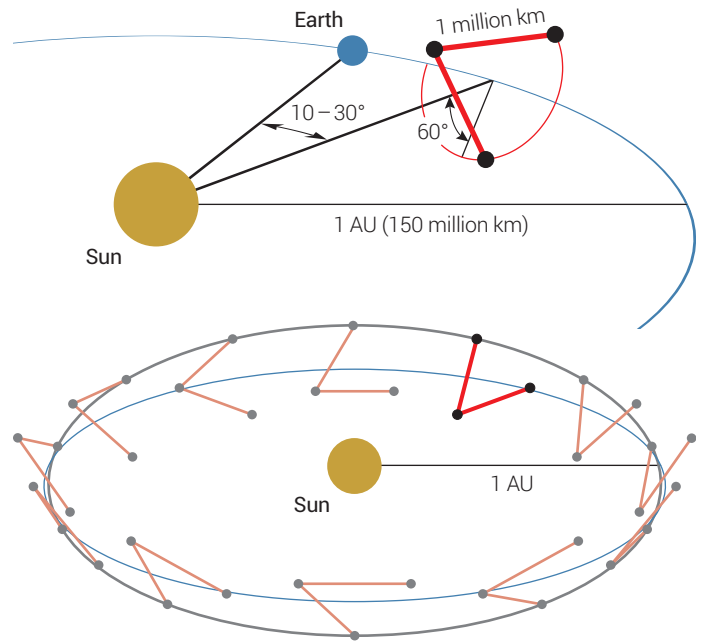

Figure 10: eLISA Orbits. The three eLISA-NGO spacecraft follow the Earth as an almost stiff triangle, purely due to celestial mechanics.

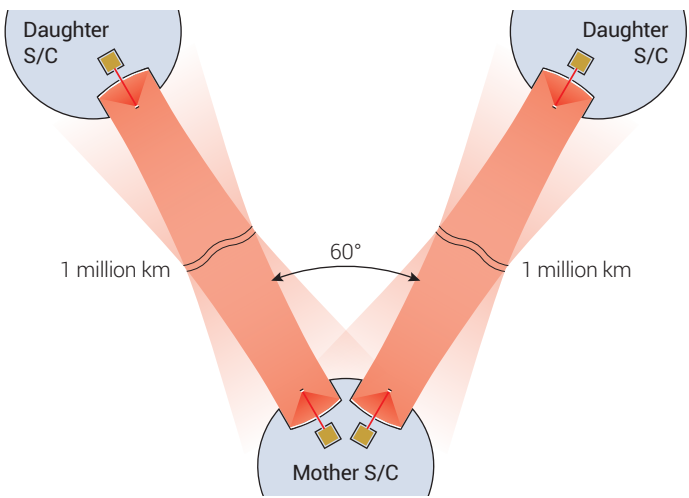

Figure 9: eLISA configuration (not to scale). One mother and two daughter spacecraft exchanging laser light form a two-arm Michelson interferometer. There are four identical payloads, one at the end of each arm, as shown in Figure 11.

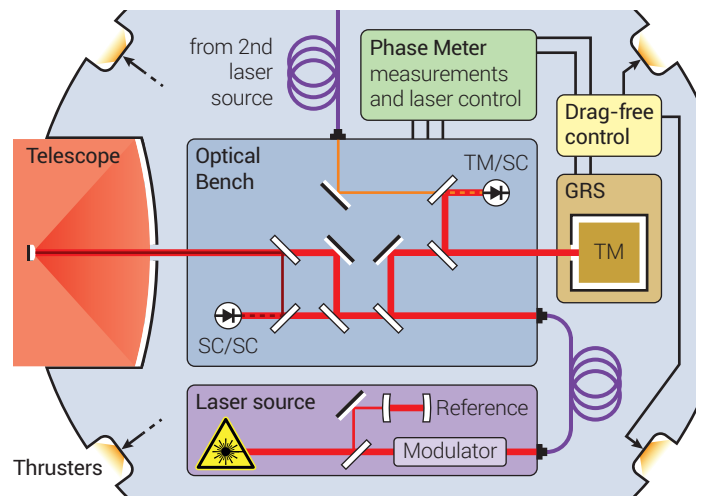

Figure 11: eLISA payload. Each payload unit contains a 20 cm telescope, the test mass enclosed inside the Gravitational Reference Sensor (GRS) and an optical bench hosting the interferometers. (Auxiliary reference interferometer omitted for clarity, see [15] for details.)

## Drag-free control

The spacecraft are actively controlled to remain centred on the test masses along the interferometric axes, without applying forces on the test masses along these axes. This 'drag-free control' around the shielded geodesic reference test masses uses the local interferometry measurement as a control signal for an array of micro-Newton spacecraft thrusters, with the residual spacecraft jitter reaching the nm/√Hz level. These thrusters also control the spacecraft angular alignment to the distant spacecraft by detecting the laser beam wavefront with 'differential wavefront sensing' with nrad/√Hz precision. Other degrees of freedom are controlled with electrostatic test mass suspensions. The only remaining degree of freedom is then the opening angle between the arms at the master spacecraft, which varies smoothly by roughly 1.5° over the year, and can be compensated for either by moving the two optical assemblies against each other or by a steering mirror on the optical bench.

The test masses are 46 mm cubes, made from a dense non-magnetic Au-Pt alloy and shielded by the GRS. The GRS core is a housing of electrodes, at several mm separation from the test mass, used for nm/√Hz precision capacitive sensing and nN-level electrostatic force actuation on all non-interferometric degrees of freedom. The GRS also includes fibres for UV light injection for photoelectric discharge of the test mass, and a caging mechanism for protecting the test mass during launch and then releasing it in orbit. The GRS technology is direct heritage from LISA Pathfinder.

## Sensitivity

The strain sensitivity (shown in Figure 12) corresponds to the noise spectrum of the instrument.

At low frequencies, it is dominated by residual acceleration noise of 3 fm s$^{-2}$/√Hz per test mass. Above about 5 mHz, arm length measurement noise dominates, for which 12 pm/√Hz are allocated, out of which 7.4 pm/√Hz are quantum mechanical photon shot noise. At the highest frequencies, the sensitivity decreases again since multiple wavelengths of the gravitational wave fit into the arms, causing partial cancellation of the signal.

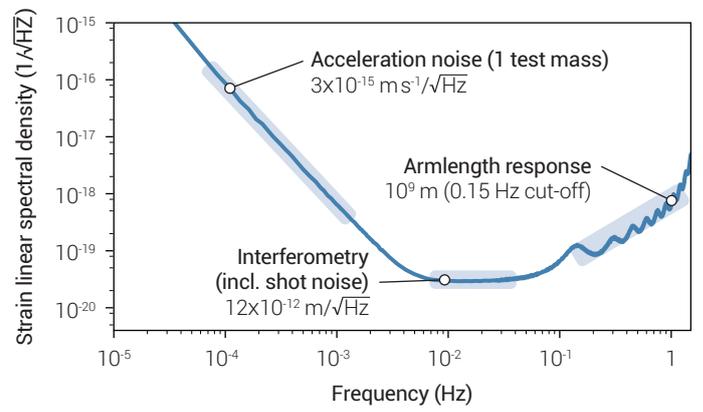

Figure 12: **Time, sky and polarisation averaged eLISA sensitivity.** The noise spectrum (strain sensitivity) is plotted as a linear spectral density.

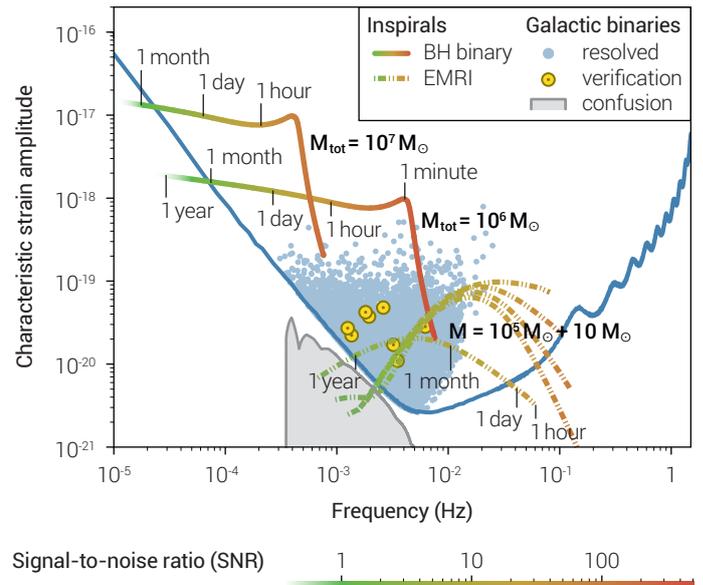

Figure 13: **Examples of gravitational wave astrophysical sources in the frequency range of eLISA, compared with the sensitivity curve of eLISA.** The data is plotted in terms of unitless 'characteristic strain amplitude'. That is the strain resolvable in a single cycle for a broadband source, or the linear spectral density of Figure 12 times √frequency, or times √ of number cycles spent by the source at a given frequency during the observation for monochromatic sources. The tracks of two massive black hole binaries, located at $z = 3$ with equal masses ($10^7$ and $10^6\,M_\odot$), are shown. The source frequency (and SNR) increases with time, and the remaining time before the plunge is indicated on the tracks. An equivalent plot is shown for an EMRI source at 200 Mpc, with 5 harmonic frequencies evolving simultaneously. Several thousand galactic binaries, with SNRs above 7, will be resolved after 1 year of observation. Some binary systems are already known, and will serve as verification signals. Millions of other binaries result in a 'confusion noise' that varies over the year. The average level is represented as grey shaded area.

A unique feature of the *eLISA* interferometry is the virtual elimination of the effects of laser frequency noise. Stabilisation to a reference cavity built into the payload is not enough to suppress it completely. The remaining noise is removed by 'Time-Delay Interferometry' (TDI) [146], which synthesises a virtual balanced arm length interferometer in postprocessing. This requires knowledge of the absolute arm lengths to roughly 1 m accuracy, measured via an auxiliary ranging phase modulation imposed on the laser beams. A second modulation is used to measure and remove noise caused by timing jitter of the Analogue to Digital Convertor sampling clocks in the phasemeters [147].

## Data Analysis

The data analysis algorithms for extracting the gravitational wave signals from the data and estimating their parameters were developed within the Mock LISA data challenge program. This program was successfully conducted from 2006 until 2011 and a new round of *eLISA* data challenges is currently being prepared. During the four rounds of the previous data challenges methods were developed for detecting gravitational wave signals from spinning massive black hole binaries, EMRIs, the population of galactic binaries, cusps formed on cosmic strings, as well as stochastic gravitational wave signals. The summary of each challenge can be found in [148–150]. The results show that we can always successfully detect and resolve gravitational wave signals, disentangling multiple sources of the same kind and of different kinds from the tens of millions of sources simultaneously present in the simulated data. In addition, the recovered parameters are always consistent with the true values to within the expected statistical uncertainty.

## Technology status

All critical technologies for *eLISA* have been under intense development for more than 15 years, and today all are available in Europe, including the phasemeter. The interferometry with million km arms cannot be directly tested on ground, but it is being studied by scaled experiments and simulations. For the *eLISA* GRS, local interferometry, and the core of the drag-free and test mass control, LISA Pathfinder has allowed early identification and resolution of both technological development challenges and performance questions (see Figure 14). The GRS force noise budget has been largely verified at the level of LISA Pathfinder, and in some respects for *eLISA*, by torsion pendulum testing on the ground [151]. Additionally, tests with the LISA Pathfinder interferometer have allowed ground verification of the local displacement measurement across the LISA Pathfinder band [152]. The 2015 Pathfinder flight represents a final verification and in-orbit commissioning of these systems and much of the *eLISA* metrology capability.

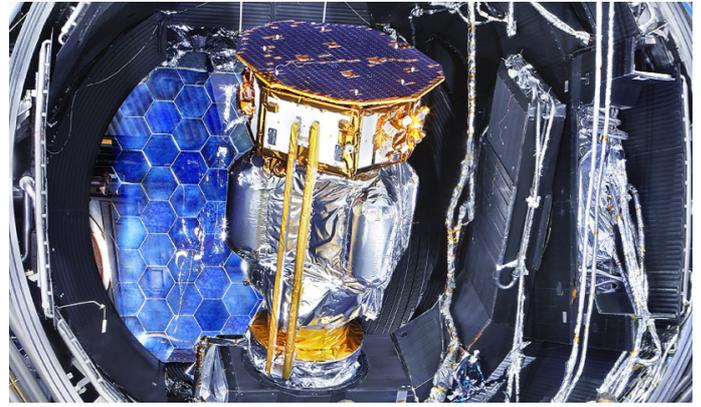

**Figure 14: LISA Pathfinder.** Much of the flight hardware for LISA Pathfinder is already tested and integrated on the spacecraft. In the last couple of years, the integrated system has been put through various test campaigns, from vibration and shock tests, to system level closed-loop communication and interface tests. The image shows the LISA Pathfinder satellite mounted on its propulsion module and situated in a space simulator in preparation for the Transit Orbit Thermal Test campaign. The space simulator aims to reproduce the thermal and vacuum environment in space. The bright light source at the back of the large vacuum tank simulates the sun allowing the spacecraft to be powered via its solar panel during the tests. The spacecraft has also undergone another test campaign in the same space simulator during which a Thermal Optical Qualification Model of the LTP (the science payload) was integrated. This second test campaign aimed to simulate the environment expected at Lagrange point 1, with the temperature in the space simulator being cycled between the minimum and maximum predicted values. This On-Station Thermal campaign included some performance tests of the optical metrology system, the results of which showed that the performance of the system is well within the requirements.

The ESA evaluation in the L1 process showed the NGO mission concept to be both technically feasible and compatible with the L2 cost target. It fits, with margin, into an Ariane V launch vehicle, though other launch scenarios (like the one studied for L1) are possible. There is strong interest from international partners, such as the US and China, to participate and contribute to *eLISA*, in which case a generous budget margin and/or enhanced science capability would be available. With or without international partners, Europe has the chance to take the lead in this revolutionary new science. ∎

---

## Science Objectives

Through the detection and observation of gravitational waves:

- Trace the formation, growth, and merger history of massive black holes
- Explore stellar populations and dynamics in galactic nuclei
- Test General Relativity with observations
- Probe new physics and cosmology
- Survey compact stellar-mass binaries and study the structure of the Galaxy

### Event Rates and Event Numbers

| | |
|---|---|
| Frequency band | $1 \times 10^{-4}$ Hz to 1 Hz, ($3 \times 10^{-5}$ Hz to 1 Hz as a goal) |
| Massive black hole mergers | $10\,\text{yr}^{-1}$ to $100\,\text{yr}^{-1}$ |
| Extreme mass ratio inspirals | $5\,\text{yr}^{-1}$ to $50\,\text{yr}^{-1}$ |
| Galactic Binaries | ~ 3000 resolvable out of a total of ~ $30 \times 10^6$ in the *eLISA* band |

# V. SCIENTIFIC LANDSCAPE OF 2028

The science capabilities of the *eLISA* mission have been described in earlier sections. *eLISA* will pioneer gravitational wave observations in the rich frequency band around 1 mHz. In this section we examine this science return in the likely context of the L2 launch date of 2028. Given the predicted state of knowledge in 2028, we ask what unique contributions *eLISA* will make to our likely understanding of fundamental physics and astronomy at that time.

Naturally, science is not predictable, and the most interesting discoveries between now and 2028 will be the ones we cannot predict! But planned projects already hint at where the frontiers of science will be when *eLISA* operates. For example, massive progress can be expected in transient astronomy. Telescopes like LSST and the Square Kilometre Array (SKA)[153] are likely to identify new systems that flare up irregularly or only once, and there is a good chance that some of these will be associated with gravitational wave signals. As another example, extremely large telescopes (EELT, TMT, GMT) and large space telescopes (JWST) will be observing (proto-)galaxies at unprecedentedly high redshifts, at which *eLISA* will simultaneously observe individual merging black hole systems. As well as providing a wealth of information that will make it easier to identify the gravitational wave sources, the expected progress in all kinds of electromagnetic astronomy will sharpen the need for complementary gravitational wave observations of the unseen Universe.

## Gravitational wave science by 2028

By 2028, gravitational wave astronomy will be well-established through ground-based observations operating at 10 Hz and above, and pulsar timing arrays (PTAs) at nHz frequencies. The huge frequency gap between them will be completely unexplored until *eLISA* is launched (see Figure 15).

The ground-based network of advanced interferometric detectors (three LIGO detectors, VIRGO[154], and the Kamioka Gravitational wave Detector, KAGRA[155]) will have observed inspiralling binaries up to around $100\,M_\odot$ and measured the population statistics. Some, or all, of these detectors will have been further enhanced in sensitivity. It is possible that the third-generation Einstein Telescope (ET) will have come into operation by 2028[156], further extending the volume of space in which these signals can be detected. At the other end of the mass spectrum, PTAs[157] will have detected a stochastic background due to many overlapping signals from supermassive black hole binaries with masses over $10^9\,M_\odot$, and they may have identified a few individual merger events. The background will help determine the mass function of supermassive black holes at the high-mass end, but it will not constrain the mass function for the much more common $10^6\,M_\odot$ black holes that inhabit the centres of typical galaxies and are ac-cessible to *eLISA*. Ground-based gravitational wave observations are unlikely to constrain the existence and population statistics of the so-far elusive intermediate-mass black holes, although optical and X-ray observations might have done so by 2028. Besides making high-sensitivity observations of individual systems, *eLISA* will characterise the population statistics of black holes in the centres of galaxies, of intermediate mass black holes, and of the early black holes that eventually grew into the supermassive holes we see today.

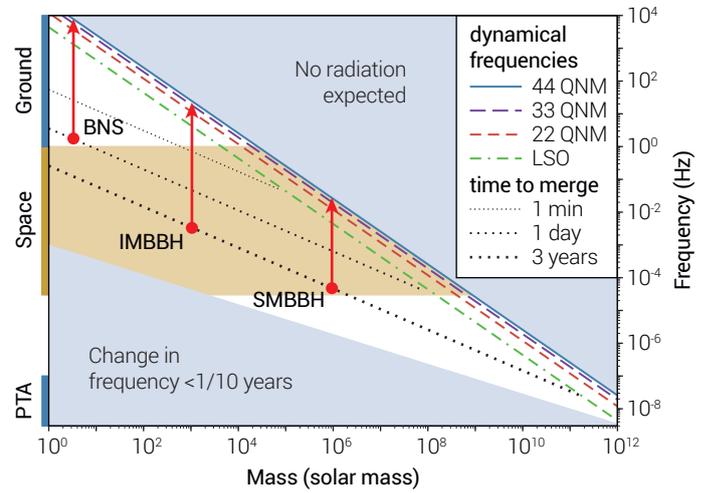

Figure 15: **Measurement capabilities of ET, eLISA, and PTAs.** The horizontal axis is the total mass of the binary system. The vertical axis is the frequency of the gravitational waves. The left side shows the frequency bands of the different instruments. The grey shaded region at the top is inaccessible because no system of a given mass can radiate at such high frequencies. The shaded region at the bottom is less interesting, because the chirp mass cannot be measured in an observation lasting less than 10 years, so the mass and distance of the source cannot be independently determined. Sloping dotted lines show the three-year, one-day, and one-minute time-to-merger lines. Sloping dashed lines are relevant dynamical frequencies: last stable orbit and the frequencies of ringdown modes of the merged black hole. Vertical lines indicate evolutionary tracks of systems of various masses as their orbits shrink and they move to higher frequencies. There is some overlap in the mass-range of sources that can be studied between ground and space detectors, but whether this translates into real science depends on the instruments' sensitivities and the source populations. Among the most interesting, would be the observation of an intermediate-mass binary black hole (IMBBH) system (if such systems exist) of around $1000\,M_\odot$ during inspiral and coalescence with both eLISA and ET. This track is shown in the diagram as (IMBBH). With such a source at redshift 0.5, both instruments might have an SNR around 20 (see Figure 16), but most of the science-including the direction to the source-would come from the eLISA measurement. Additionally, tracks for binary neutron stars (BNS) and super massive binary black holes (SMBBHs) are shown.

By 2028, theoretical advances and predictable improvements in computer power will have made it possible to compute the complex waveforms expected from EMRIs and supermassive black hole binaries with high precision. This will allow searches in *eLISA* data to approach the optimum sensitivity of matched filtering, and it will make tests of General Relativity using these signals optimally sensitive.

## eLISA and fundamental science in 2028

One of the signature goals of *eLISA* is to test gravitation theory, and it seems unlikely that any other method will achieve the sensitivity of *eLISA* to deviations of strong-field gravity by 2028. Unlike ground-based instruments, *eLISA* will have sufficient sensitivity to be able to notice small corrections to Einstein gravity, and possibly to recognise unexpected signals that could indicate new phenomena.

By observing the long-duration waveforms from EMRI events, *eLISA* will map with exquisite accuracy the geometry of supermassive black holes, and will detect or limit extra scalar gravity-type fields. The MICROSCOPE (Micro-Satellite à traînée Compensée pour l'Observation du Principe d'Equivalence) mission [158] will, by 2028, have improved our limits on the violation of the equivalence principle or could, of course, have measured a violation, which would make *eLISA*'s strong-field observations even more urgently needed. X-ray and other electromagnetic observatories may measure the spins of a number of black holes, but *eLISA*'s ability to follow EMRI and merger signals through to the formation of the final horizon will not have been duplicated, nor will its ability to identify naked singularities or other exotic objects (such as boson stars or gravastars), if they exist.

By 2028 we will know much more about the large-scale Universe: in particular, about the nature of dark energy from the upcoming optical surveys dedicated to probing the large-scale structure. However, many questions will have remained open concerning the early Universe. From Planck and balloon flights of CMB instruments, we may know how much primordial gravitational radiation was produced at the end of inflation, which will help to pin down the actual inflationary scenario. However, there exists no means other than gravitational waves to probe the period in the evolution of the Universe ranging from reheating after inflation until big bang nucleosynthesis. Through the detection of gravitational waves, *eLISA* can gather information on the state of the Universe at much earlier epochs than those directly probed by any other cosmological observation. Gravitational waves are the next messengers to probe the very early Universe. They reach beyond the cosmic microwave background: *eLISA* has access to a fundamental frequency/energy-scale window, that of TeV. This scale is presently our boundary of knowledge in fundamental particle physics; new physics is therefore expected to emerge around that scale.

It is unclear how much progress will have been made by 2028 in understanding fundamental particle physics. The LHC will start probing new physics after its first upgrade in 2015 (reaching the scale of 14 TeV); another upgrade by a factor of 10 is expected around 2022, which will take the experiment through to 2030. It is difficult to foresee what the LHC will reveal about the nature of the Higgs, of dark matter particles, supersymmetry, extra dimensions, and so on. High-sensitivity *eLISA* observations may be crucial in providing clues here, since they explore the relevant energy scales in a completely unique way. The information contained in gravitational waves from the early Universe is complementary to, and independent of, the one accessible by particle accelerators. The presence of a first order phase transition at the TeV scale, the presence of cosmic (super-)strings in the Universe, the properties of low-energy inflationary reheating, even the nature of the quantum vacuum state before inflation began (which could be different from the standard Quantum Field Theory nature in loop quantum gravity), are some of the fundamental issues that will still be open in 2028, and to which *eLISA* might provide some answers.

### eLISA and late-time cosmology in 2028

By 2028 our understanding of the way cosmological structures formed will have been dramatically improved by high-redshift observations of QSOs and protogalaxies from missions like JWST [159], EUCLID [160] and the Wide-Field Infrared Survey Telescope (WFIRST) [161], and by the Atacama Large Millimeter/submillimeter Array (ALMA) [162] on the ground. These observations may well have constrained the supermassive black hole mass spectrum from a few times $10^{10}\,M_\odot$, or even higher, down to around $10^7\,M_\odot$, but probably not into the main *eLISA* range of $10^4 – 10^6\,M_\odot$, especially at $z > 2$. *eLISA* observations will fill this gap and also provide a check on selection effects and other systematics of the electromagnetic observations. By being able to measure the mass and spins of massive black holes as a function of redshift out to $z = 20$, *eLISA* will allow us to greatly improve models of how supermassive black holes grow so quickly, so as to be in place at $z \sim 7$. We will additionally learn what roles accretion and mergers play in the growth of all massive black holes. *eLISA* observations of mergers of $10^4 – 10^5\,M_\odot$ black holes out to $z = 20$ (if they exist) can provide a strict test of the amount of growth by merger expected in these models.

### eLISA and massive black holes in 2028

In the next few years, eROSITA will study tidal disruptions of stars out to redshifts of $z \sim 1$ and will look for massive black holes [163], although in the high-mass regime compared to *eLISA*. *eLISA* has extraordinary sensitivity to massive black holes in the mass-range characteristic of most galactic-core black holes (see Figure 16). Gravitational wave detectors like *eLISA* are inherently all-sky monitors: always on and having a nearly $4\pi$ steradian field of view. They naturally complement other surveys and monitoring instruments operating at the same time, like LSST [164], SKA, neutrino detectors, gamma-ray and X-ray monitors. The massive black hole mergers detected by *eLISA* out to modest redshifts ($z = 5 – 10$) could well be visible to SKA and LSST as transients in the same region of the sky. The identification of 5 to 10 counterparts during a 2 year *eLISA* mission would not be surprising. These might then be followed up by large collecting area telescopes like TMT, GMT, and EELT, providing an unprecedented view of the conditions around two merging massive black holes. Interestingly, the advent of observing with detectors like aLIGO is leading to the development of networks of optical telescopes for multimessenger astronomy. These are designed to follow up gravitational wave triggers and find associated transient phenomena. These systems may usefully supplement LSST in picking out *eLISA* counterparts.

*eLISA* has very good sensitivity to as many as 4 or 5 ringdown frequencies of newly formed black holes (see Figure 16). This 'black hole spectroscopy' will allow *eLISA* to address important questions that, by 2028, will probably not yet have been answered. Most important will be to find evidence for naked singularities or other exotic objects: The ringdown frequencies make it possible to determine the mass and spin of the final black hole, and will be very different for any of the proposed alternatives. In addition, the ringdown modes show in detail how a dynamical black hole behaves; not even ET, if operational at that time, will have the sensitivity to make this kind of detailed study of strong gravity.

### eLISA and the astrophysics of stars and the Galaxy in 2028

*eLISA* will perform, for the first time, a complete census of very compact binary systems throughout the Galaxy. Thousands of white-dwarf binaries are expected, along with binaries involving neutron stars and black holes in various combinations. GAIA's catalogue will still, in 2028, be the principal optical reference for these observations, and we can expect dozens or more binaries in that catalogue to be observed by *eLISA*. *eLISA* will identify many more, the nearest of which can then be followed up with JWST if still operating, and E-ELT. These observations will lead to improved understanding of interactions, mass transfer, and double white dwarfs as supernova progenitors. By 2028, aLIGO and partners will have good statistics on the population of relativistic compact binaries out to Gpc distances, and *eLISA*'s complete census of that population in the Galaxy will allow us to compare our Galaxy with the cosmological norm, a comparison that is very difficult to make with any other stellar population, revealing much about the history of our Galaxy. If binaries of $100\,M_\odot$ black holes exist, then *eLISA* and ET could make joint observations of a few merging systems with comparable sensitivity, improving on the angular positions which ET could measure alone.

The Event Horizon Telescope (EHT) should test general relativity and probe the horizon of Sgr A*, the massive black hole at the centre of the Milky Way [165]. EHT should also explore the shape and properties of the accretion flow onto Sgr A* and measure its spin [166–167].

Within the next few years, GRAVITY [168] may be able to observe the orbits around Sgr A* of currently unknown stars with periods of just 1 – 2 years, test General Relativity, and infer the hidden distribution of the dark population of objects around Sgr A*, thereby studying the different models for mass-segregation in galactic nuclei [61, 66, 169–170]. Such studies should provide more accurate estimates of the event rate for EMRIs in the *eLISA* band. ∎

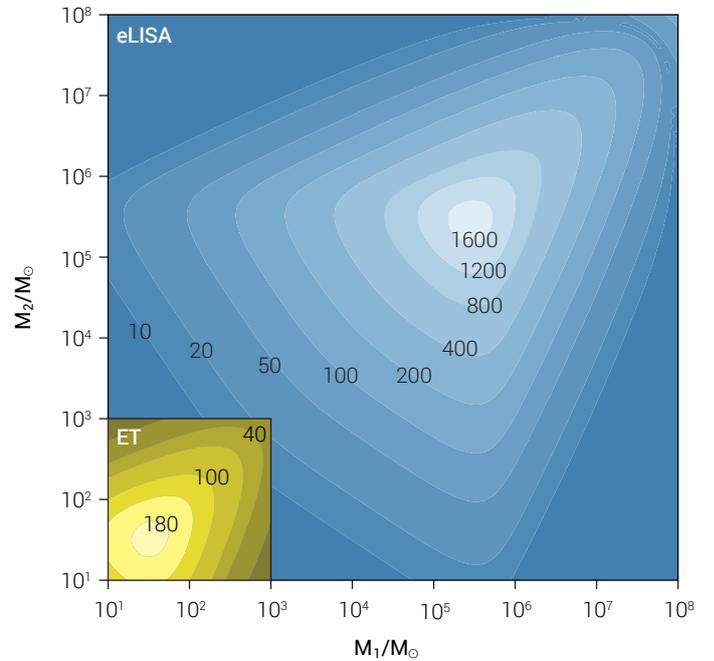

Figure 16: **SNR contours of eLISA and ET for binary black holes at redshift 0.5.** The axes are the masses (in solar masses) of the two components of the binary. The main contour plot is for eLISA, and the box in the lower left shows the contours for ET, on the same axes. The two observatories overlap for sources at this distance only for binaries with comparable masses, around 1000 $M_\odot$. For smaller distances, the SNR scales as $1/r$; for larger redshifts the SNR also scales inversely with the luminosity distance and the masses redshift as well, so that the axis scales show $(1+z)\,M$, where $M$ is the intrinsic (rest-frame) mass. The SNR values have been calculated using only the inspiral part of the waveform; they underestimate the true sensitivity by omitting the merger and ringdown radiation. This is particularly important at the higher mass end for each detector.

# CONCLUSION

In summary, by 2028 our understanding of the Universe will have been dramatically improved by advanced observations of electromagnetic radiation. Adding a low-frequency gravitational wave observatory will add a new sense to our perception of the Universe. Gravitational waves will allow us to 'hear' a Universe otherwise invisible with light.

*eLISA* will be the first ever mission to survey the entire Universe with gravitational waves. It will allow us to investigate the formation of binary systems in the Milky Way, detect the guaranteed signals from the verification binaries, study the history of the Universe out to redshifts of order 20, test gravity in the dynamical strong-field regime, and probe the early Univserse at the TeV energy scale. *eLISA* will play a unique role in the scientific landscape of 2028.

The NGO mission studied by ESA for the L1 mission selection serves as a strawman mission concept that is capable of matching the science requirements for *eLISA*. It has been evaluated by ESA as both technically feasible and compatible with the L2 cost target. ∎

*Links to the original documents can be found at*
*http://elisascience.org/references*

**On the cover page:** Gravitational waves generated by the white dwarf pair RX J0806.3+1527 (artist's impression) in front of the Milky Way's Galactic Centre (wide spectrum composite image) with black hole Sagittarius A*.
*Credit: NASA, ESA, SSC, CXC, STScI, and AEI.*